# A CRYPTOSYSTEM BASED ON HILBERT MATRIX USING CIPHER BLOCK CHAINING MODE

Penmetsa V Krishna Raja [#1], A S N Chakravarthy [*2], Prof. P S Avadhani [#3]

[#]1Research Scholar, JNTUK, Kakinada, Andhra Pradesh, India
[*]2Associate Professor , Department of CSE& IT, Sri Aditya Engineering College, Surampalem, , Andhra Pradesh, India
[#]3Professor, Dept. of CS & SE, Andhra University, Visakhapatnam , Andhra Pradesh, India

*Abstract*— **Cryptography is the science of using mathematics to encrypt and decrypt data. Cryptography enables you to store sensitive information or transmit it across insecure networks so that it cannot be read by anyone except the intended recipient. While cryptography is the science of securing data, cryptanalysis is the science of analyzing and breaking secure communication. Classical cryptanalysis involves an interesting combination of analytical reasoning, application of mathematical tools and pattern finding. The objectives of the proposed work are to propose a new cryptographic method based on the special matrix called the Hilbert matrix for authentication and confidentiality and to propose a model for confidentiality and authentication using a combination of symmetric and public cryptosystems. Further, it is extended to shared key cryptosystems with the concept of digital enveloping using a session key.**

**In the present work an algorithm for shared key encryption is developed using Hilbert matrix cryptosystem. In this the block chaining modes of operation have been used to tackle the issues of confusion and diffusion.**

## I. INTRODUCTION

Cryptosystems can be categorized as four types, namely, Identity based, group key based; lattice based and shared key cryptosystems. In secret sharing a secret (password, key, trade secret,...) is used as a seed to generate a number of distinct secrets, and the pieces are distributed so that some subset of the recipients can jointly authenticate themselves and use the secret information without learning what it is. Secret sharing is also called secret splitting or key splitting.

We want to share N secrets among M people so that any M < N of them (M of N) can regenerate the original information, but no smaller group up to M − 1 can do so. There are mathematical problems of this type, such as the number of points needed to identify a polynomial of a certain degree or the number of intersecting hyper planes needed to specify a point ad used in Blakley's scheme [1].

Shared-secret encryption algorithms use a single secret key to encrypt and decrypt data. You must secure the key from access by unauthorized agents because any party that has the key can use it to decrypt data. Shared-secret encryption is also referred to as symmetric encryption because the same key is used for encryption and decryption. Shared-secret encryption algorithms are extremely fast (compared to public-key algorithms) and are well suited for performing cryptographic transformations on large streams of data.

The disadvantage of shared-secret encryption is that it presumes two parties have agreed on a key and IV and communicated their values. Also, the key must be kept secret from unauthorized users. Because of these problems, shared-secret encryption is often used in conjunction with public-key encryption to privately communicate the values of the key and IV.

## II. METHODOLOGY

First the existing symmetric cryptosystems, namely, DES, AES, IDEA, Blow Fish and RC5 have been studied. The advantages and limitations of the number of rounds, the key size and the block size in each of them have been observed. In the second phase, the public key cryptosystems and the concepts of session key and the concept of the digital enveloping are studied. The public key systems include RSA, Diffie-Hellman Key exchange and NTRU. The underlying hard problems of number theory, abstract algebra and elliptic curves, namely, the factorization problem and the discrete logarithm problems are studied. While studying these, it is observed that Hilbert matrices have special properties which can be utilized for developing a symmetric cryptosystem. They are then analyzed and found suitable. A symmetric cryptosystem is designed, developed and implemented.

### A. Special Matrices

In this section some special matrices will be presented whose properties will be useful in developing some cryptosystems. For example, the set of all n x n non-singular matrices form a non-abelian group under the operation of multiplication of matrices.

Presently, there are many known matrix based cryptosystems, they are not secure with the present day computing environment as they all have only polynomial time complexity. However, there are some points to ponder in these methods which can be applied in some other form. Some of them are As any encryption/decryption algorithm is computationally intensive it is better to have those matrices whose inverses must have all integer entries. The matrices used as the key need to be highly unstable making it difficult to compute the inverse.

If the size n of the key matrix K is unknown then the computation of K-1 is virtually impossible. This makes it interesting because there are some special matrices with these properties which can be better utilized to design an encryption and decryption scheme. The proposed work is based on this idea. The following types of the matrices are of special importance.

a) Vandermonde's matrix [2][3], b) Combinatorial matrix, c) Cauchy's matrix and d) Hilbert matrix which is a special case of Cauchy matrix.

1) The Vandermonde's matrix $V(x) = (a_{ij})_{n \times n}$ is a square matrix where $a_{ij} = x_j^i$ Observe that this matrix has the terms of a geometric progression in each row [3].

2) The Combinatorial matrix $C(x) = (a_{ij})_{n \times n}$ is a square matrix with

$$a_{ij} = y + \delta_{ij} x$$

where

$$\delta_{ij} = \begin{cases} 0 & \text{if } i \neq j \\ 1 & \text{if } i = j \end{cases}$$

3) Cauchy's matrix is given by $Cu(x, y) = (a_{ij})_{n \times n}$ is a square matrix where $a_{ij} = 1/(x_i + y_i)$

4) Finally the Hllbert matrix is a special case of the Cauchy matrix with $x_i + y_j = i + j - 1$.

Thus the Hilbert Matrix of order n is explicitly given by

$$\begin{bmatrix} 1 & \frac{1}{2} & \frac{1}{3} & \cdots & \frac{1}{n} \\ \frac{1}{2} & \frac{1}{3} & \cdots & \cdots & \frac{1}{n+1} \\ \frac{1}{3} & \frac{1}{4} & \cdots & \cdots & \frac{1}{n+2} \\ \frac{1}{4} & \vdots & \cdots & \cdots & \cdots \\ \vdots & \vdots & \cdots & \cdots & \cdots \\ \frac{1}{n} & \vdots & \cdots & \cdots & \frac{1}{2n-1} \end{bmatrix}$$

*B. Properties of Special Matrices*

As the proposed algorithms use Hilbert matrices, the properties of Hilbert matrices are of much importance[4]. Some of the useful ones are listed below.

(i) The Hilbert matrix is a special case of the Cauchy matrix.
(ii) It is symmetric and positive definite.
(iii) It is also totally positive meaning the determinant of every sub matrix is positive.
(iv) The determinant can be expressed in closed form, as a special case of the Cauchy determinant. The determinant of the n × n Hilbert matrix is

$$\left[ \prod_{1 \leq i < j \leq n} (i-j)^2 \right] \Big/ \left[ \prod_{1 \leq i,j \leq n} (i+j-1) \right]$$

The determinant of the Hilbert matrix is the reciprocal of an integer.

(v)
(vi) The Hilbert matrix is invertible for any order.
(vii) The entries of the inverse matrix are all integers.

Although all the above properties are interesting it is particularly useful for us the last two properties and a proof is given in the following.

(viii) The Inverse of the Hilbert matrix H is equal to

$$= \left[ \prod_{k=0}^{n-1} (i+k)(j+k) \right] \Big/ \left[ (i+j-1) \prod_{\substack{k=1 \\ k \neq i}}^{n} (i-k) \prod_{\substack{k=1 \\ k \neq j}}^{n} (j-k) \right]$$

(ix) The Determinant of the combinatorial matrix is

$$x^{n-1}(x+ny)$$

(ix) The Determinant of Vandermonde's matrix is

$$\prod_{1 \leq j \leq n} x_j \prod_{1 \leq i \leq j \leq n} (x_j - x_i)$$

(x) The Determinant of Cauchy's matrix is

$$\prod_{1 \leq i \leq j \leq n} (x_j - x_i)(y_i - y_i) \Big/ \prod_{1 \leq i,j \leq n} (x_i + y_i)$$

(xi) The Inverse of a combinatorial matrix is a combinatorial matrix is a combinatorial matrix with the entries

$$b_{ij} = \left( -y + \delta_{ij}(x+ny) \right) \Big/ x(x+ny)$$

(xii) The inverse of Vadermonde's Matrix is given by

$$b_{ij} = \left( \sum_{\substack{1 \leq k_1 < \ldots < k_{n-j} \leq n \\ k_1, \ldots, k_{n-j} \neq i}} (-1)^{j-1} x_{k_1} \cdots x_{k_{n-j}} \right) \Big/ x_i \prod_{\substack{1 \leq k \leq n \\ k \neq i}} (x_k - x_i)$$

(xiii) One can observe that the complicated sum in the numerator is just the coefficient of xj-i in the

polynomial

$$(x_1-x)\cdots\cdots(x_n-x)/(x_i-x).$$

(xiv) The inverse of Cauchy's Matrix is given by

$$b_{ij}=\left(\prod_{1\le k\le n}(x_j+y_k)(x_k+y_i)\right)\bigg/\left((x_j+y_i)\left(\prod_{\substack{1\le k\le n\\k\ne j}}(x_j-x_k)\right)\left(\prod_{\substack{1\le k\le n\\k\ne i}}(y_i-y_k)\right)\right)$$

Although there are many special properties for each of these matrices our interest is to specific about Hilbert Matrix. Hilbert Matrix is very unstable [5] and hence it can be used in security systems.

### C. Existing Cryptosystems Using Matrices

It has been proposed in the literature that there are some cryptosystems based on matrices, [4][6]. However, all the general matrix operations are in the class P. The best-Known multiple-letter encryption cipher is the Playfair cipher which treats diagrams in the plaintext as single units and translates these units into cipher text diagrams. The Playfair algorithm is based on the use of a 26 letters constructed using a keyword as a 5 x 5 matrix. The key word is chosen with no repetitions. The matrix is constructed by filling in the letters of the keyword from left to right and from top to bottom, and then filling in the remainder of the matrix with the remaining letters in alphabetic order. The letters I and J count as one letter. Plaintext is encrypted two letters at a time, according to the following rules.

Repeating plaintext letters that are in the same pair are separated with a filler letter, such as x. Two plaintext letters that fall in the same row of the matrix are each replaced by the letter to the right, with the first element of the row circularly following the last.

Two plaintext letters that fall in the same column are each replaced by the letter beneath, with the top element of the column circularly following the last.

Otherwise, each plaintext letter in a pair is replaced by the letter that lies in its own row and the column occupied by the other plaintext letter [2].

Despite the level of confidence in its security, the playfair cipher is relatively easy to break because it still leaves much of the structure of the plaintext language intact. A few hundred letters of cipher text are generally sufficient.

**Hill Cipher:**

Another interesting multi-letter cipher based on matrix theory is the Hill cipher [7]. The encryption algorithm takes m successive plaintext letters and substitutes for them m cipher text letters. The substitution is determined by m linear equations in which each character is assigned a numerical value with a = 0, b = 1…….z =25. Then the equation will be

$$C = K^P \bmod 26$$

Where C and P are column vectors of length n, representing the plaintext and cipher text and K is an n x n invertible matrix, representing the encryption key. Here the matrices K and P are multiplied and then the operation 'mod 26' is performed on them to get the matrix C. The elements of C are then converted as a, b, c….. as per their values. Decryption is carried out using the matrix equation

$$P = K^{-1} C \bmod 26$$

where $K^{-1}$ is the matrix inverse of K.

The problem with $K^{-1}$ is it may not yield integer values. In fact more often the inverses of integer matrices end up with their elements as real numbers. This slightly complicates the above equation as we have to convert the corresponding entities into integers. There are many ways proposed in the literature for that [7]. As with Playfair cipher, the strength of the Hill cipher is that it completely hides single letter frequencies. Indeed with this, the use of a larger matrix hides more frequency information. Thus a 3 x 3 Hill cipher hides not only single letter but also two letter frequency information and so on. Although the Hill cipher is strong against a cipher text only attack, it can be broken with a known plaintext attack.

The properties of Hilbert matrix are exploited in developing a cryptosystem by [8]. The main property being used here is that the inverse of any Hilbert matrix are integers. However, it still suffers from the language restrictions. Hence to have good confusion and diffusion, the concept of block chaining modes of operation are used for the algorithm presented in [8]. Here, again the instability along with the property that all the entries in its inverse are integers plays an important role.

### D. Encryption And Decryption

In this section, the encryption and decryption algorithm is proposed in [1][8] is first presented and Cipher Block Chaining Modes are being used on that. In the algorithm proposed by Suresh, the size n of the Hilbert Matrix is the secret key. The size n will be sufficiently large so that the plain text can be taken as an n x 1 column matrix. As the key n is known to both sender and the receiver, the actual Hilbert matrix and also its inverse are also known to both of them. At the sender's side, the plaintext column is multiplied by the Hilbert Matrix to get another n x 1 column matrix of cipher text. However, there are problems in keeping the size n secret because the size is dependent on the plain text. Hence, first the size is taken as secret key and the size of the plain text block is taken as smaller than n. Let the size of the plaintext is m. Now another (n-m) x 1 secret column K is also transmitted securely to the other side. Thus, instead of the plaintext being an nx1 column, it will be an mx1 column and the secret column K of size (n-m) x 1 is appended. The resultant column of length n will be encrypted with the Hilbert matrix and sent to the other side. However, the language characteristics may still yield some security breaches. Hence cipher block chaining modes are proposed.

On the receiver's side, in the decryption process, as n is already known to the receiver, the inverse of the Hilbert

Matrix of order n is also known. The received cipher which is an n x 1 column matrix, is then multiplied inverse Hilbert matrix to get the n x 1 column containing the plain text column and the secret column K is also known to the receiver, the plain text is known have more complexity, n is taken as a large prime n The algorithms for encryption and decryption are given Here the plaintext P is taken as a column matrix of order 1 and n is a prime number much larger than m. Her public but n is a secret known only to the sender a receiver. Further a string K of length n-m is also cho both sender and receiver and kept secret.

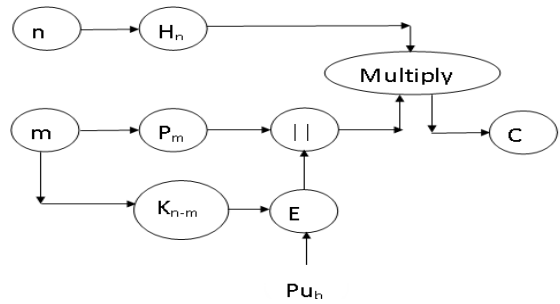

Fig. 1 Encryption using Hilbert matrix

### III. CRYPTOSYSTEM BASED ON HILBERT MATRIX

The key generation, encryption and decryption of Hilbert matrix cryptosystem are presented here. Here m is block size and known to all.

*A. Key Creation*

Step 1: Choose a large prime number n and communic securely between both the parties.

*B. Encryption*

Step 1: Let M be the plain text.
Step 2: Write the plaintext message as an m x 1 column matrix
Step 3: Choose a secret string K and encrypt with the pul key of the receiver.
Step 4: Convert the result into a (n-m) x 1 column matrix
Step 5: Append this column matrix to make it an n x 1 column.
Step 6: Multiply the n x n Hilbert Matrix with this n x 1 column. The resultant is the cipher text.
Step 7: Transmit the cipher text to the receiver.

*C. Decryption*

After receiving the cipher text by the receiver, he will decrypt it using the following algorithm.

Step 1: As the size n of the Hilbert Matrix is known, its inverse matrix is also known, which is an n x n matrix . Further, the cipher text is also known as an n x 1 column matrix.
Step 2: Multiply the inverse Hilbert matrix with the n x 1 cipher column matrix. The resultant contains the plain text and also the encrypted secret value of K
Step 3: Detach this and get the plain text. Observe that we need not decrypt the encrypted secret value as we are not interested in it. The algorithm is illustrated in figure 1.

In this diagram P is the plain text, K is a secret key, H is the Hilbert matrix and C is the cipher text

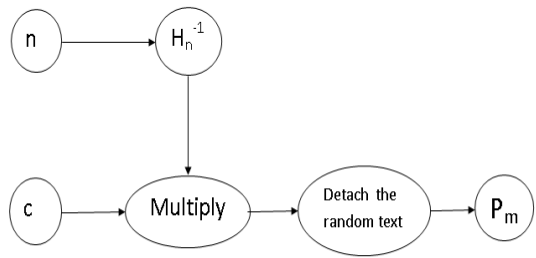

Fig. 2: Decryption using Hilbert Matrix

The difficulty in this model is that the size n and also the secret string must be sent securely. Actually, it is sufficient if we send the size m of the plain text. It is really unnecessary to send the whole of the value K. We can use the concept of the public key cryptography to solve this by sending them as session keys. The basic idea in this is to send the sizes of the Hilbert matrix and the plain text strings as a session keys to the receiver.

### IV. CIPHER BLOCK CHAINING MODES

Actually in any block cipher algorithm if a plain text block P results in the corresponding cipher text C and another plain text block P' results in the cipher text block C' and if P = P' then C will be equals to C'. Thus, in a plain text message if two blocks are same then the corresponding cipher text blocks will be same. To overcome this security defficiency, we would like a technique in which the same plain text block, if repeated, produces different ciphertext blocks. This is called the cipher block chaining (CBC) mode. In this scheme, the input to the encryption algorithm is the XOR of the current plaintext block and the preceding cipher text block; the same key is used for each block. In effect, we have chained together the processing of the sequence of plaintext blocks. The input to the encryption function for each plain text block bears no fixed relationship to the plaintext block. Therefore, repeating patterns of blocks are not exposed.

For decryption, each cipher block is passed through the decryption algorithm. The result is XORed with the peceding ciphertext block to produce the plaintext block. Mathematically it can be represented as follows.

$$C_j = E_k[C_{j-1} \text{ XOR } P_j] \text{ and } P_j = C_{j-1} \text{ XOR } D_k[C_j]$$

To produce the first block of ciphertext, an initialization vector (IV) is XORed with the first block of plaintext. On decryption, the IV is XORed with the output of the decryption algorithm to recover the first block of plaintext.

The IV must be known to both the sender and receiver. For maximum security, the IV should be protected as well as the key. One reason for protecting the IV is as follows: If an opponent is able to fool the receiver into using a different value for IV, then the opponent is able to invert selected bits in the first block of plaintext. The block chaining modes of operation is depicted in figure 3 below.

In addition to its use to achieve confidentiality, the CBC mode can be used for authentication. These block chaining modes of operation is applied to the encryption algorithms discussed in the preceding sections of this chapter.

## V. ALGORITHM USING BLOCK CHAINING MODES OF OPERATION

The algorithm presented in the above section, based on the Hilbert matrix suffers from the following problem s. If a the plain text P is divided into blocks, say, P = [P1, P2,...., Pt] and the corresponding cipher text blocks C is divided into C = [C1, C2, C3,....., Ct] then if $P_i = P_j$ then $C_i = C_j$. This implies that, if for any reason, two cipher text blocks are same then a successful attack to get the corresponding plain text block reveals the other plain text block also without any extra effort. This is avoided by using the block chaining modes of operation, the algorithm of which is presented here. We assume that the key as well as an initial value IV of the same size of the plain text block are securely communicated between the sender and the receiver.

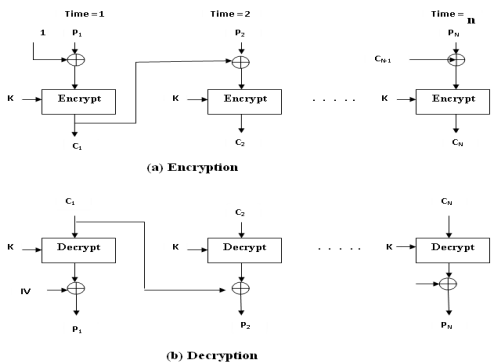

Fig.3 Block chaining modes operation

### A. Encryption
Step 1: Let M be the plain text.

Step 2: Write the plaintext message as t blocks of suitable size as M = [ $M_1$, $M_2$, $M_3$, ......., $M_t$]
Step 3: Choose the initial value IV of the same size of each of the Mi which is already sent to the receiver securely.
Step 4: Put $C_0$ = IV
Step 5: For i = 0 to t do
{
  Step 5.1: Compute $Y_i = C_i$ XOR $P_{i+1}$
  Step 5.2: Compute $C_i = E_k(Y_i)$ where $E_k(Y_i)$ denotes the encryption of $Y_i$ using the algorithm III(B) above.
}
Step 6: Combine the resultant to get C = [ $C_1$, $C_2$, ......., $C_t$] which is the cipher text sent to the receiver.
Step 7: Transmit the cipher text C to the receiver.

### B. Decryption
After receiving the cipher text C by the receiver, he will decrypt it using the following algorithm.
Step 1: Divide C into t blocks as C = [ C1, C2, ......., Ct]
Step 2: For i = 0 to t do

{
  Step 2.1: Compute $Y_i = D_k(C_i)$ where $D_k(C_i)$ denotes the de encryption of Ci using the algorithm III(C) above.
  Step 2.2 : Compute $Y_i$ XOR $C_i = P_i$ with $C_0$ = IV.
}
Step 3: Combine the resultant to get P = [ $P_1$, $P_2$, ......., $P_t$] which is the plaintext.

It can easily be observed that even if two cipher text block are same their corresponding plain text block will be total different. This is because each block of the plaintext is tied up the cipher text of the previous block. This helps us handle redundancy in blocks successfully.

## VI. CONCLUSIONS

In the present work a new cryptosystem based on Hilbert Matrices, which is an improvement of has been proposed. The idea behind choosing the Hilbert matrices is that they are unstable and have integer inverses . These properties make them interesting and they are listed below.
1. Whatever may be the order, Hilbert matrices are invertible.
2. The inverse of a Hilbert matrix will have all its entries integers.
3. If the order is known then the inverse can be easily found and it is very difficult to find the inverse if the order is unknown.
4. Direct computation of the inverse of the Hilbert matrix leads to more round off errors due to its unstable nature.

All these properties proved to be an advantage for the designing a cryptosystem based on Hilbert matrix. As the size of the Hilbert matrix is kept secret (known only to sender and receiver) because of the instability and the point 4 above, it is difficult and practically impossible for anyone to retrieve the message without knowing n. Further, as the inverse has closed form solution with all its entries being integers, the round of errors can be avoided. As n needs to be kept secret and the plain text size is supposed to be public, we need to have another string to be augmented to the plain text. This is also kept secret and this secret string K is used as a session key and also can be used for authentication.

## VII. FUTURE DIRECTIONS

The share key algorithm presented here can be extended for authentication in a shared environment further in the place of secret random garbage value, any key or equivalent can be transmitted securely. These can be used in applications like copyright protection and digital water marking because of its simplicity and ease of use. This can be extended to other models of security like the group key communication.

## Authors

[1] Penmetsa V Krishna Raja received his M.Tech (CST) from A.U, Visakhapatnam, Andhra Pradesh, India. He is a research scholar under the supervision of Prof.P.S.Avadhani. His research areas include Network Security, Cryptography, Intrusion Detection, Neural networks.

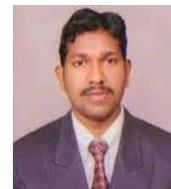

[2]A.S.N Chakravarthy received his M.Tech (CSE) from JNTU, Anantapur , Andhra Pradesh, India. Presently he is working as an Associate Professor in Dept. Of Computer Science and Engineering in Sri Aditya Engineering College, SuramPalem, AP, India. His research areas include Network Security, Cryptography, Intrusion Detection, Neural networks.

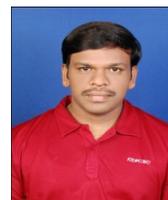

[3] Prof. P.S.Avadhani did his Masters Degree and PhD from IIT, Kanpur. He is presently working as Professor in Dept. of Computer Science and Systems Engineering in Andhra University college of Engg., in Visakhapatnam. He has more than 50 papers published in various National / International journals and conferences. His research areas include Cryptography, Data Security, Algorithms, and Computer Graphics, Digital Forensics and Cyber Security.

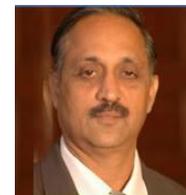